# Virtual spectrometer for stable radicals' vibrations. 2. Graphene molecules

Elena F. Sheka and Nadezhda A. Popova

Peoples' Friendship University of Russia (RUDN University), 117198 Moscow, Russia

*Corresponding author: E.F.Sheka

**Abstract.** The article presents the result of an extended *in silico* experiment on computational vibrational spectroscopy of graphene molecules performed using the virtual vibrational spectrometer UHF VVS, previously proposed in the first part of the study [1]. Since *in vitro* spectroscopy of such individual molecules is practically impossible, the obtained spectra give the first idea of what these spectra could be. A large set of studied molecules allows drawing the first conclusions about the specific features of the vibrational dynamics of the substances, caused by their radical structure and spin features of the ground electronic state.

## 1. Introduction

The idea of a virtual spectrometer turned out to be very attractive and productive for putting things in order in the field of computational spectroscopy. Any device is usually designed to carry out a large number of measurements. Certainly, these measurements are performed under conditions limited by the technical characteristics of the device. Whatever these characteristics are, there will always be reasons for their inconsistency with some specific requirement of real experiment. This circumstance is largely compensated by a comparative analysis of the data obtained from the device, which greatly increases their reliability. The technical characteristics of a virtual spectrometer are determined by the computational softwares, on which it is based. In the case of vibrational spectroscopy, the programs must allow solving the dynamical problem of an object using the methods of quantum theory. Thus, in the first realized virtual multi-frequency spectrometer (VMS) [1], the calculations are carried out using available CCSD(T) and DFT codes, supplemented with VPT2 approximation concerning dynamical problem. In the second case of UHF VVS [2], the calculations are performed in the Hartree-Fock approximation using its restricted (RHF) and unrestricted (UHF) version. The result of the five-year use of the VMS [3] was to clarify the requirements for basic programs that would provide a fairly close agreement of the calculated vibrational and electronic spectra of small molecules. In the second case, the main attention was directed to the calculation of the vibrational spectra of molecular radicals of large sizes, carried out using the semi-empirical UHF approximation. Within the framework of the basic computational program, corrections to the elements of the force matrix are not provided, as a result of which a accurate fitting of the computed spectrum to experimental data could not be provided and, thus, was not a target. The calculations were concentrated on revealing general trends accompanying the object radical nature under the condition of a close similarity of general structural images of virtual and experimental spectra. A comparative analysis of the RHF-UHF pairs of IR and Raman spectra with experimental data allowed establishing a tight correspondence of the general structure of the spectra and made it possible to reveal a number of trends characteristic of the vibrational spectra of molecules of different classes, which was shown by the example of polyacenes and fullerenes. So, in the course of this study, the influence of the radical nature of molecules on their vibrational spectra was established. A natural change in the structure of the vibrational IR and Raman spectra of polyacene caused by increasing the number of benzenoid rings was revealed. An answer was obtained to the question of the origin of 'silent' modes in the vibrational spectra of $C_{60}$ and $C_{70}$ fullerenes. The main methodical conclusion of the study was the establishment of the fact that the UHF approximation, which is the basis of the technical characteristics of the UHF VVS, does make it possible to reproduce IR and Raman spectra of molecular radicals quite reliably. This conclusion lays the foundation for the next step in the UHF VVS use in the current study applied to graphene molecules.

Graphene molecules (GMs) present a particular class of compounds related to nanosize graphene sheets, both bare and framed by a necklace of heteroatoms in due course of the polyderivatization of different kinds. Molecules of all the types are radicals [4], among which the former are much stronger. Bare GMs do not exist either in Nature or in a laboratory, which does not prevent them from being favorite objects of the virtual graphenics [5]. In contrast, framed GMs are stable radicals and are largely known as, say, basic structure units (BSUs) of largely distributed $sp^2$ amorphous carbons in the Nature and mass produced in laboratories and industry [6]. Existing in amorphous solid media, the molecules vary in shape and chemical composition due to which experimental vibrational spectra of the relevant amorphous solids [6, 7] give only an averaged presentation of the dynamics of their BSUs. Recent studies of the structure and chemical composition of a set of amorphous carbons [8], supplemented by their neutron and photon scattering spectra as well as IR absorption, are a good basis for putting a question what are these BSUs and what is vibrational image of graphene molecules in general, the molecules that do not exist as discrete individuals. It is time to turn to virtual spectroscopy to get answer. This article presents the first results on the way obtained using the UHF VVS.

## 2. Graphene molecules to be considered

Figure 1 presents a set of graphene molecules, virtual IR and Raman spectra of which were exhibited with the UHF VVS. The first raw in the figure is occupied with bare molecules **I**, **II**, and **III** related to two monolayer graphene rectangular sheets (5x5) and (9x9) characterized by 5 and 9 benzenoid units along the armchair and zigzag edges, respectively. Molecule **III** correspond to two-layer (5x5) structure. The choice of model **I** is provided by two reasons: 1) the (5x5)NGr molecule was the basic model for computational consideration of a large spectrum of properties related to spin chemical physics of graphene [9]; 2) this molecule is commensurate with the predominant part of the BSUs of the studied $sp^2$ amorphous carbons [6-8]. Subjecting the latter to a virtual polyderivation allowed suggesting a set of reliable molecular models of the units. The second raw in Fig. 1 is given to three hydrogen-stimulated poliderivatives of molecule **I** (mono (**IV**) and poly (**V**) hydrogenated graphenes and polymethylated graphene (**VI**)). The third raw presents three poliderivatives related to oxygenated graphene molecules (graphene molecules framed with oxygene atoms (**VII**), as well as with hydrohyl (**VIII**) and carboxyl (**IX**) units, respectively). The last forth raw involves the BSU models suggested for the studied amorphous carbons (shungite carbon (**X**), anthraxolite (**XI**), anthracite (**XII**) and carbon black (**XIII**)) [6-8], IR and Raman spectra of which present the pool of experimental data to be compared with virtual ones. Criterion parameters, identifying radical state of the molecules, are listed in Table 1. All the calculations discussed below were performed using CLUSTER-Z1 software [10] implementing semi-empirical UHF codes. An AM1 version was explored. The spectra chemical attribution was discussed on the basis on the general frequencies (GFs) concept [11, 12], selected set of which, related to the studied molecules, is listed in Table 2. Due to usual upshifting of vibrational frequencies provided with UHF calculations (see discussion in [1]), a necessary interrelation of the experimental and UHF VVS frequencies data, attributed to the same GFs, is suggested in column 2 of the table. This interrelation was confirmed in the course of a comparative analysis of experimental and virtual UHF VVS vibrational spectra of pentacene and fullerenes $C_{60}$ and $C_{70}$ [1].

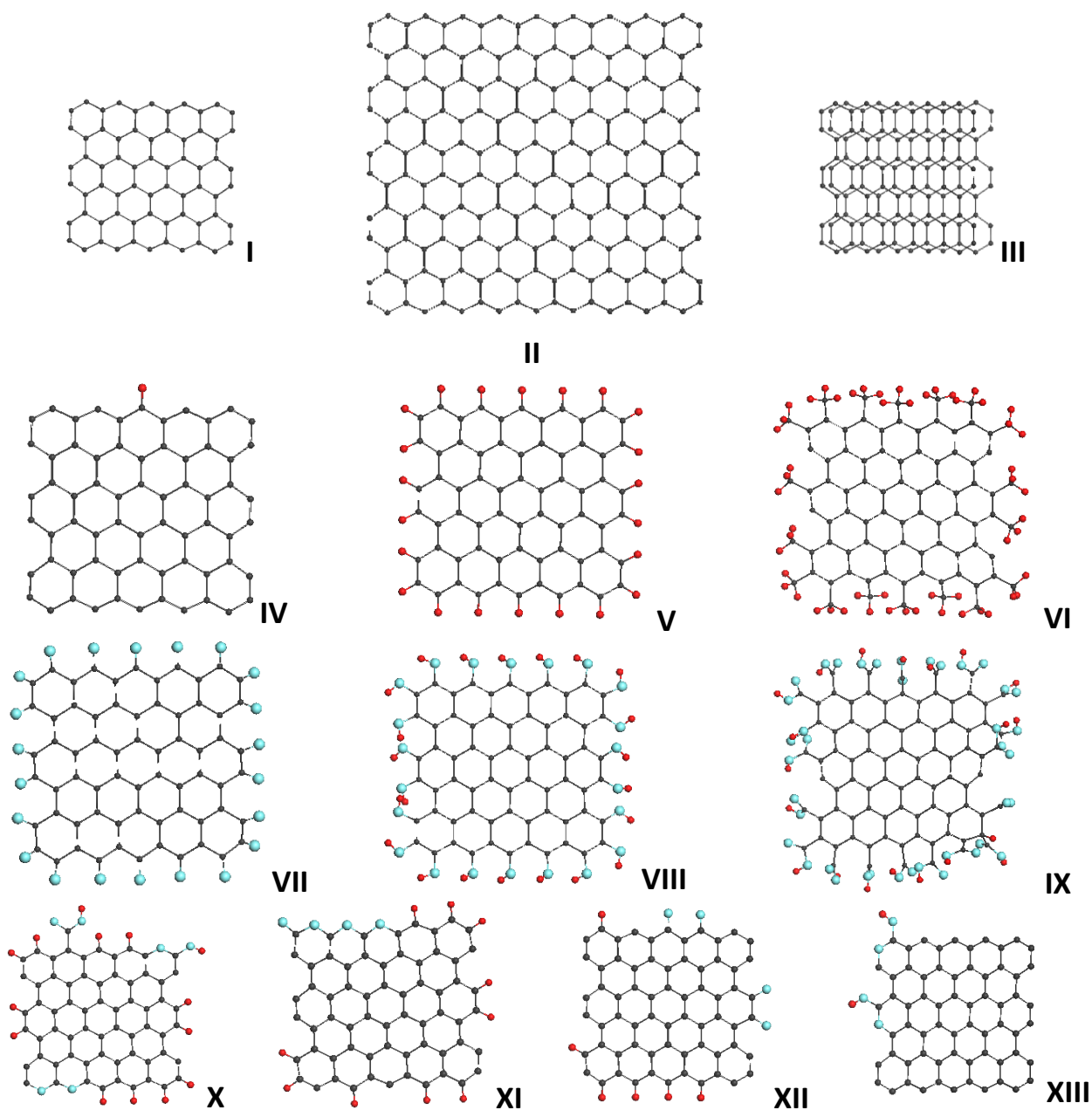

**Figure 1**. Equilibrium structures of the graphene molecules under study: **I** – (5x5)NGr; **II** – (9x9)NGr; **III** – two layers of (5x5)NGr; **IV – VI** graphene hydrides; **VII – IX** graphene oxides; **X – XIII** models of the basic structure unite of *sp²* amorphous carbons: anthracite (**X**), anthraxolite (**XI**), shungite carbon (**XII**), and carbon black (**XIII**) (see text for details).

## 3. Bare graphene molecules[1]

The IR and Raman UHF VVS spectra of the basic GM **I** of the current study is presented in Fig. 2 alongside with the spectra of fullerenes $C_{60}$ and $C_{70}$. In terms of the number of atoms, it is in the middle between the fullerenes. The same can be said about its spectra as well. As seen in the figure, in case of IR spectra, the main changes in spectra concern the region of C=C stretchings over 1200

---

[1] Here and therein in the paper all digital wavenumber notations are related to the UHF VVS data (see comments to Table 2).

cm⁻¹ where the spectrum of the GM is closer to that of $C_{70}$. The absence of fullerene tangential modes at ~550 cm⁻¹ in the graphene spectrum is evidently connected with the flat molecule structure in contrast to either sphere- or ovaloid-like shapes of fullerenes. As discussed earlier [1],

Table 1. Identifying criterion parameters of the odd electron correlation in graphene molecules[1]

| Molecule number | Chemical formula | Point symmetry | Number of atoms | Spin multiplicity $\Delta N_{\alpha\beta}$[2] | $N_D$, e⁻ | $\Delta \hat{S}_U^2$ |
|---|---|---|---|---|---|---|
| I | $C_{66}$ | $D_{2h}$ | 66 | triplet | 34,2 | 18,1 |
| II | $C_{99}$ | $C_2$ | 190 | Singlet | 76,2 | 38,1 |
| III | 2 $C_{66}$ | $C_s$ | 132 | triplet | 67,6 | 34,8 |
| IV | $C_{66}H$ | $C_{2v}$ | 67 | doublet | 33 | 16,8 |
| V | $C_{66}H_{22}$ | $D_{2h}$ | 88 | triplet | 14,2 | 8,1 |
| VI | $C_{66}(CH_3)_{18}$ | $C_1$ | 138 | singlet | 20 | 10 |
| VII | $C_{66}O_{22}$ | $C_1$ | 88 | singlet | 16 | 8 |
| VIII | $C_{66}(OH)_{22}$ | $C_1$ | 110 |  | 110 | 110 |
| IX | $C_{66}(COOH)_{22}$ | $C_1$ | 146 | singlet | 20 | 10 |
| X | $C_{66}O_6H_{14}$ | $C_1$ | 86 | triplet | 20,6 | 11,3 |
| XI | $C_{63}O_4H_{10}$ | $C_1$ | 77 | triplet | 21,6 | 11,8 |
| XII | $C_{66}O_4H_6$ | $C_1$ | 76 | triplet | 25 | 13,5 |
| XIII | $C_{64}O_4H_2$ | $C_1$ | 70 | singlet | 29,2 | 14,6 |

[1] Identifying criteria are discussed in [2].
[2] Formal spin multiplicity is determined by the difference $\Delta N_{\alpha\beta} = N_\alpha - N_\beta$, $N_\alpha$ and $N_\beta$ are the numbers of electrons with spins α and β, respectively

Table 2. Standard group frequencies of aromatic molecules required for the fractional analysis of vibrational spectra of amorphous carbons

| Wavenumbers, cm⁻¹ | | Group frequencies[1] | | | | | |
|---|---|---|---|---|---|---|---|
| Exp. | UHF VVS | (C, C)[2] | (C, H₁)[2] | (C, CH₂)[3] | (C, CH₃)[4] | (C, O₁) and (C, OH)[5] | (C, O-C) |
| 400-700 | 600-900 | 404 δ *op* C-C-C<br>606 δ *ip* C-C-C | - | 711 ρ CH₂ | 210 *r* CH₃<br>344 δ CH₃ | 458 τ COH | 605<br>δ C-O |
| 700-1200 | 900-1400 | 707 δ C-C-C puckering,<br>993 ring breathing,<br>1010 δ C-C-C trigonal | 673<br>δ *op* in phase<br>846<br>δ *op*, C₆ libration<br>967 δ *op*<br>990<br>δ *op*, trigonal<br>1037 δ *ip*<br>1146<br>δ *ip*, trigonal<br>1178 δ *ip* | 948 ρ CH₂ | 900 ν C-CH₃<br>1041 ρ CH₃ | 960<br>ν C-O(H)<br>1158<br>δ C-O(H)<br>1284<br>ν C-O(H) | 970<br>ν C-O<br>1260<br>ν C-O |
| 1200-1600 | 1400-1800 | 1309<br>ν C-C Kekule,<br>1482 ν C-C<br>1599 ν C-C | 1350 δ *ip* in phase | 1409<br>δ internal CH₂ | 1333 δ CH₃<br>1486<br>δ internal CH₃ | 1511 ν C=O | - |

| 1600-1900 | 1800-2100 | - | - | | | 1500-1700[6] ν C=O | |
|---|---|---|---|---|---|---|---|
| 2800-3200 | 3000-3400 | - | 3056 ν C-H 3057 ν trigonal C-H 3064 ν C-H 3073 ν in phase C-H | 3114 ν CH$_2$ | 2950 ν CH$_3$ | 3400 ν OH | - |

[1] Greek symbols ν, δ, ρ, r, τ mark stretching, bending, rocking, rotational, and torsion modes, respectively.
[2] GFs notifications of fundamental vibrations of benzene molecule [13].
[3] GFs notifications of fundamental vibrations of benzyl radical [14, 15]. Hereinafter, GFs, additional to the benzene pool of vibrations, will be shown only.
[4] GFs notifications of fundamental vibrations of toluene [16, 17].
[5] GFs notifications of fundamental vibrations of p-benzosemiquinone [18].
[6] GFs notifications of fundamental vibrations of a large collection of organic molecules [11, 12].

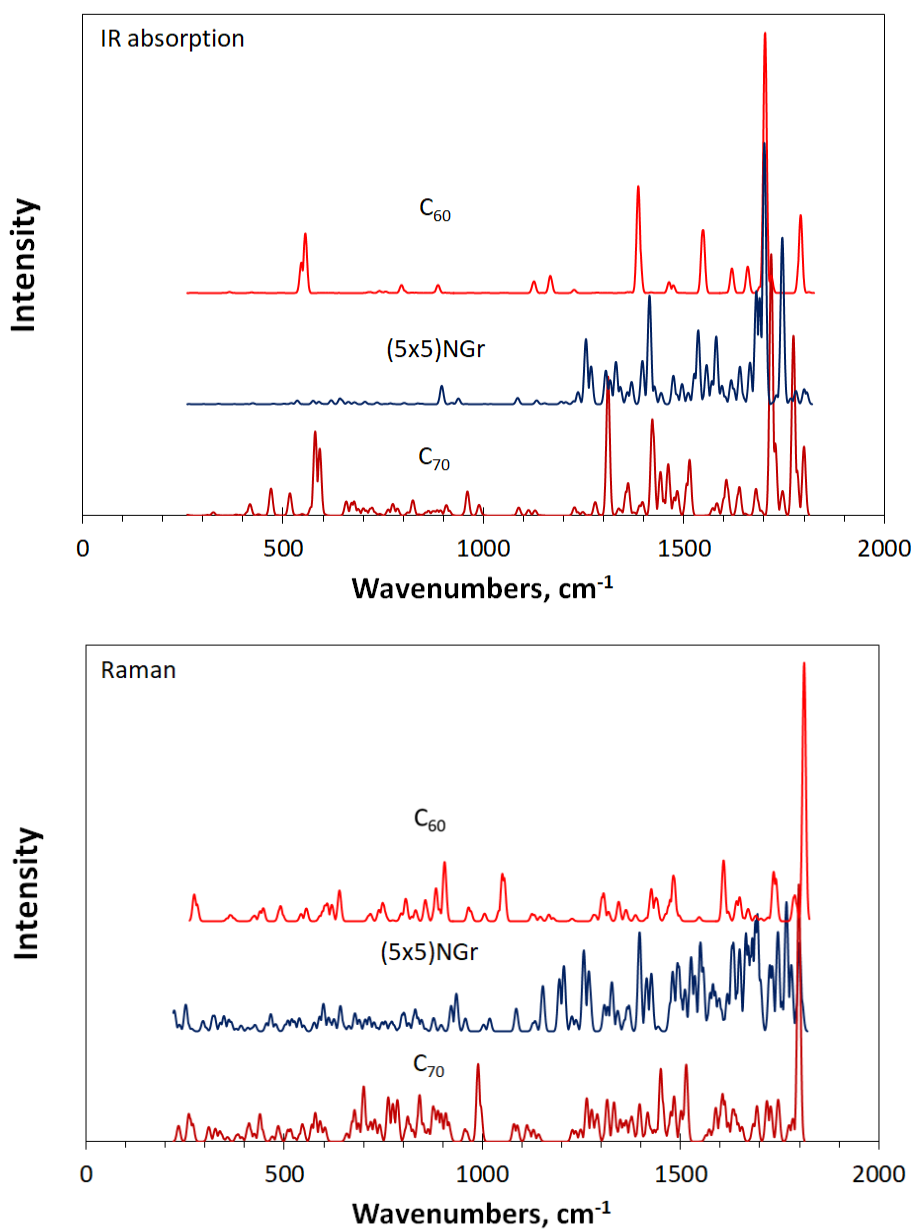

**Figure 2.** Virtual IR absorption and Raman spectra of fullerene C$_{60}$ (red), nanographene (5x5) (molecule **I**, dark blue), and fullerene C$_{70}$ (dark red). Stick-bar data are convoluted with Gaussian bandwidth of 10 cm$^{-1}$. Intensities are reported in arbitrary units, normalized per maximum values. Raman spectrum of the (5x5)NGr molecule is additionally scaled (see text).

Raman spectra of both fullerenes are of highly specific shape with strongly dominating narrow bands at ~1800 cm$^{-1}$ and extended spectra of 'silent' modes up to 200 cm$^{-1}$. The dominants are absent in the graphene spectrum thus confirming high symmetry and round molecular shapes responsible for their appearance in the case of fullerenes. Instead, once 'silent' in fullerene spectra, C=C stretchings in the region of 1200-1800 cm$^{-1}$ become the main battlefield for the Raman spectrum of graphene. This feature is specific for graphene in all the cases, as we shall see in what follows. Since all the three molecules are configured by C=C bonds, force constants of the relevant dynamic problem are strongly dependent on the bond lengths [19]. Obviously, any redistribution over bond length is followed with a mandatory redistribution of the force constants over vibrational modes, which results in changing the vibrational spectra image. Figure 3 presents the bond length distribution of the studied three molecules. Clear difference in the plottings convincingly evidences the spectra difference in details.

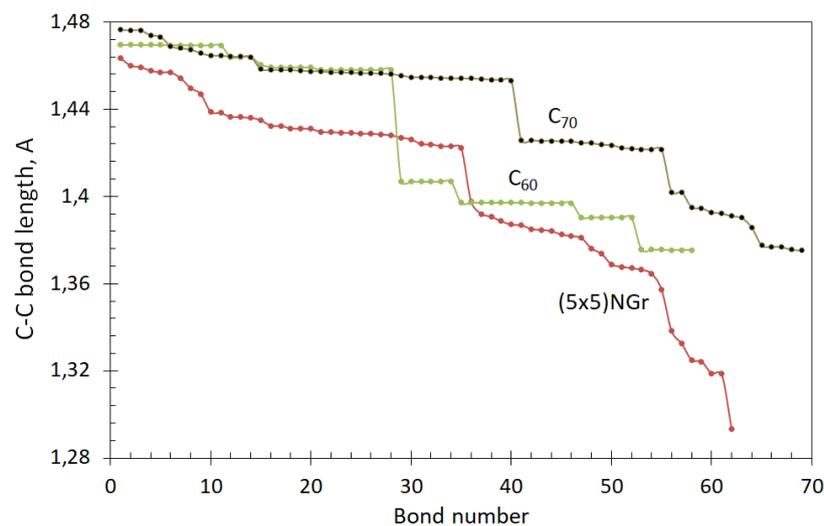

**Figure 3**. $Z \to A$ distribution of the C=C bond length.

Figure 4 exhibits transformation of the vibrational spectra of bare graphene molecules caused by either growing the molecule size or layering the pristine one. As seen in the figure, only C=C stretchings contribute into the spectra shape. Existence of spikes related to particular modes greatly suppress another modes leaving impression of a rather weal IR absorption of bare GMs. Another situation occurs in Raman spectra presented in the figure. As previously, the main action involves C=C stretchings. Important to note significant changing in the Raman spectra image of the pristine molecule when growing the molecule size (going up) and doubling the number of layers (going down). The up changing demonstrates a significant simplification of the spectrum fine structure towards the spectrum of ordered crystalline species. As found [7], the free path of the optical phonon in graphene related to C=C stretchings constitutes $L_{ph} \sim 15$ nm. The lateral size of molecule **II** is 4nm. It is still much less than $L_{ph}$, however, since the high sensitivity of the molecule Raman spectra on geometrical parameters is obvious, the germination of a crystal-like behavior can be expected before achieving $L_{ph}$. On the other hand, two-layer composition results in the spectrum resembling D-G doublet structure of nanosize graphenous materials [20]. As will be seen later, Raman spectra of none of the studied GMs do not have such a pattern, which makes us pay special attention to this finding and raise the question of the role of multilayer stacks of GMs, which are usually present in all studied graphene materials, in the formation of this characteristic doublet structure of their Raman spectra.

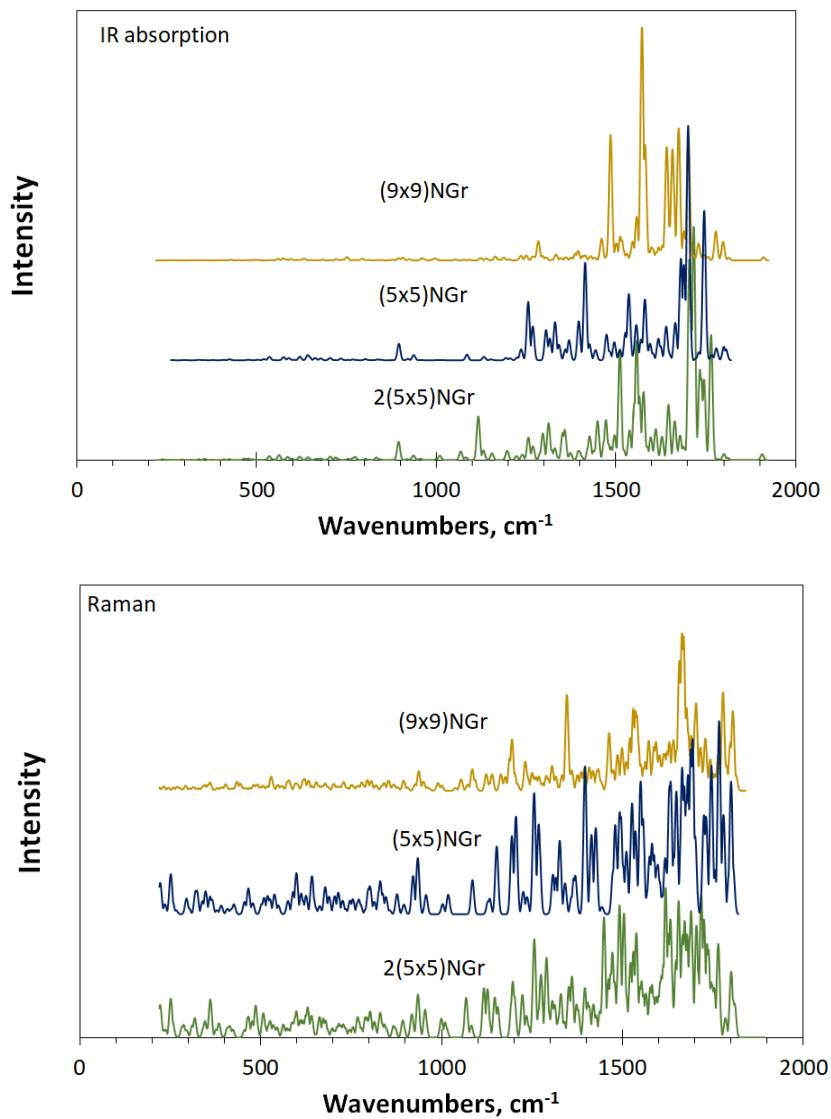

**Figure 4.** Virtual IR absorption and Raman spectra of bare graphene molecules **I** (dark blue), **II** (dark green), and **III** (dark yellow). Stick-bar data are convoluted with Gaussian bandwidth of 10 cm$^{-1}$. Intensities are reported in arbitrary units, normalized per maximum values.

## 4. Graphene molecules with hydrogen necklaces

Hydrogen is the favorite participant of carbonaceous compounds and GMs are not exception. Graphene is friendly with hydrogen, which, both in natural and industrial conditions, is one of the first elements to partially inhibit the activity of initially bare GMs. A detailed discussion of the graphene hydrogenation, both virtual and experimental, can be found in Ref. 9 and 21. The hydrogen role for natural molecules is largely discussed by the example of natural amorphous carbons [22, 23]. In this work, this process will be considered from the point of view of virtual vibrational spectroscopy using the example of artificial molecules **IV**, **V**, and **VI** and amorphous carbon models **X**, **XI**, **XII**, and **XIII** (see Table 1 and Fig. 1).

Virtual spectra of the molecules of the first set, produced with the UHF VVS, are shown in Fig. 5. As seen in the figure, the presence of hydrogen significantly influences both IR and Raman spectra of the pristine (5x5)NGr molecule, starting form one atom in the molecule necklace (1.5 at %) (see spectra of molecule **IV**). In the IR spectrum, the hydrogen presence becomes apparent in this case through the characteristic high frequency band of C-H stretchings at 3200 cm$^{-1}$ and a significant intensification of the band of benzenoid ring breathings at 1000 cm$^{-1}$, which initially was considerably weaker. In the Raman spectrum, we see the appearance of the C-H stretching band at 3200 cm$^{-1}$ and a considerable reconstruction of the spectrum of C=C vibrations, particularly, in the 1200-1500 cm$^{-1}$ region. So strong influence of one hydrogen atom on the dynamic of the molecule consisting from 66 heavy atoms is highly impressive and is observed for the first time. Evidently, the feature is connected with a particular delocalization of the molecule electrons density over the honeycomb core atoms, making any chemical attachment of the carbon core non-local [24].

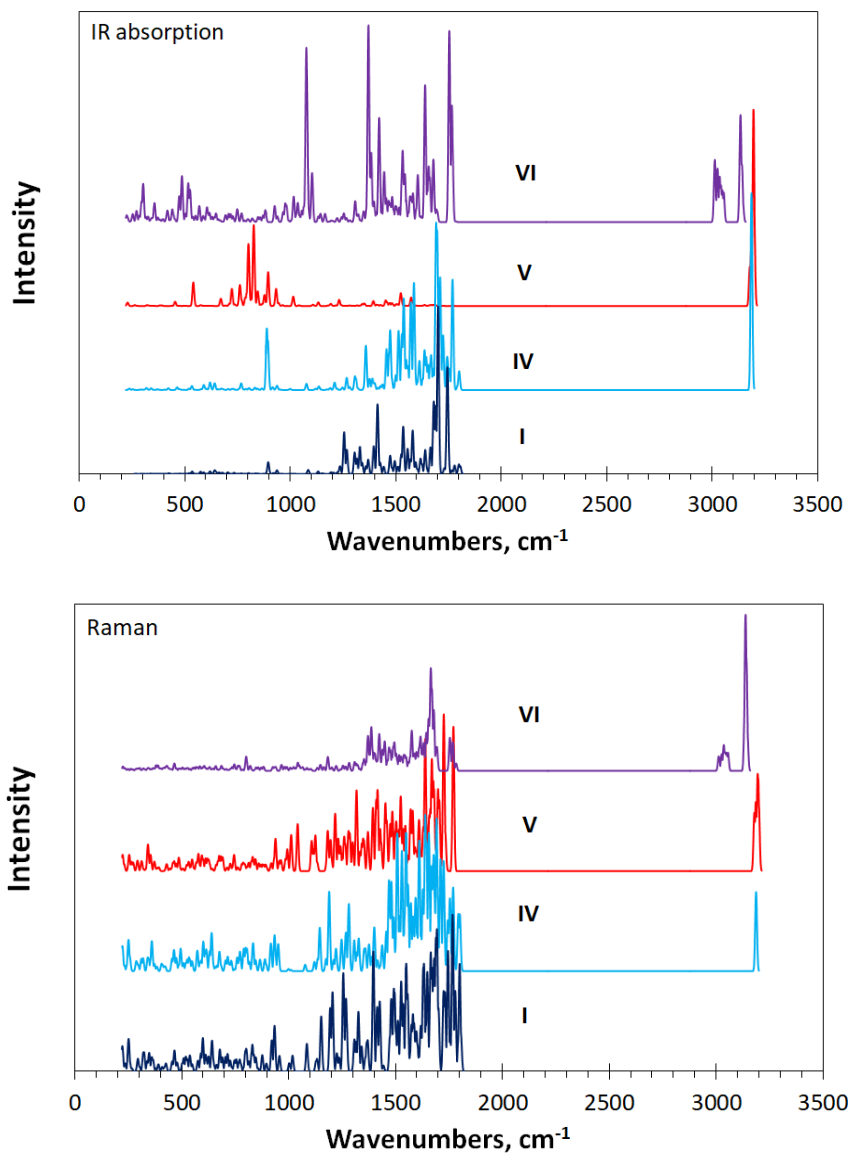

**Figure 5.** Virtual IR absorption and Raman spectra of pristine graphene molecule (**I**, dark blue) with different hydrogen necklaces consisting of one hydrogen atom (**IV**, blue), 22 hydrogen atoms (**V**, red), and 18 CH$_3$ units (**VI**, violet). Stick-bar data are convoluted with Gaussian bandwidth of 10 cm$^{-1}$. Intensities are reported in arbitrary units, normalized per maximum values.

The necklace of 22 atoms corresponds to the case (molecule **V**), when all edge atoms of the pristine molecule **I** are once determined by hydrogens (see Fig. 1). As seen in Fig. 5, the situation in the IR spectrum is drastically changed. The relative contribution of all the C=C vibrations is practically fully suppressed, while C-H bendings and C-H stretchings play the role in the regions of 700-1000 cm$^{-1}$ and 3200 cm$^{-1}$, respectively. In contrast, the molecule Raman spectrum is not changed so drastically, leaving a comparable contribution of the C-H stretchings with respect to that of molecule

**IV** and evidencing the main role of the C=C vibrations, only accompanied the changing in the relative spectral image.

Substitution of hydrogen atoms with methyl units (molecule **VI**) leads to a full reconstruction of both IR and Raman spectra of the molecule. As seen in Fig. 5, in the high frequency part of both spectra, a single bend of individual C-H stretchings in the spectra of molecule **IV** is replaced with two-band spectra of the set of $CH_3$-units C-H stretchings. The middle frequency part of the IR spectrum is filled with $C-CH_3$ stretchings as well as $C-CH_3$ and $CH_3$ internal bendings in the regions of 900-1200 $cm^{-1}$ and 1200-1800 $cm^{-1}$, respectively. Therefore, IR spectra of molecules **V** and **VI** are explicit signature of the graphene molecules vibrations related to their necklaces. The feature was earlier suggested when analyzing experimental IR spectra of amorphous carbons [6]. Unexpextedly, the same can be said about the Raman spectrum of molecule **VI** in Fig. 5. Nothing else, but a complete suppression of the contribution of C=C vibrations and, instead, visualization of the scattering on the $CH_3$ modes, can explain the observed change of the spectrum with respect to that of molecule **V**. Important to note that edge carbon atoms are fully (or almost fully in the case of molecule **VI**) terminated in both cases. Nevertheless, when hydrogens are directly attached to the carbon core in molecule **V**, they are on a distant spacing through C-C bonds in molecule **VI**. Apparently, the feature explains why C-H vibrations just dissolve in the scattering pool of the C=C modes in the first case while become dominating scatters in the second. This is an evident revealing of the difference related to the non-local and local attachments of hydrogens. The feature is one more exhibition of unique physicochemical properties of graphene molecules and deserves a further study.

As for changes in the Raman spectra of C=C vibrations of molecules **I**, **IV**, and **V** presented in Fig. 5, a reasonable explanation of the feature concerns the difference of the C=C bonds pool in the relevant cases. Actually, Fig. 6 shows the distributions of the molecules C=C bonds over their length. As seen in the figure, the observed changes are quite remarkable to be revealed in the relevant spectra.

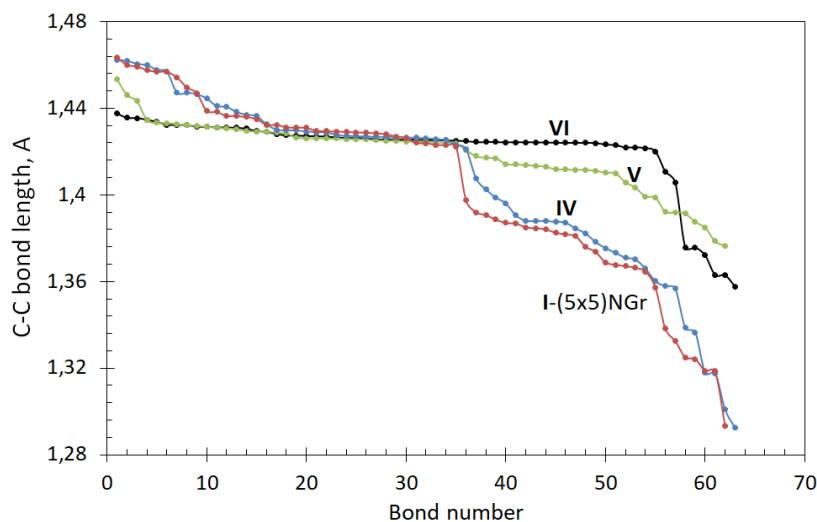

**Figure 6**. $Z \rightarrow A$ distribution of the C=C bond length in graphene hydrides (see text).

### 5. Graphene molecules with oxygen necklaces

A set of the UHF VVS produced IR and Raman spectra of molecules **VII-IX** (see Fig. 1) is presented in Fig. 7. Oxygen containing necklaces of all the molecules correspond to a complete (by O atoms, molecule **VII**, and by OH units, molecule **VIII**) or almost complete (by COOH units, molecule

**IX**) termination of the pristine (5x5)NGr molecule edge atoms. Analysis of the spectra of molecules **IV**-**VI** performed in the previous section greatly facilitate that one of the spectra in the figure. As seen in Fig. 7, IR spectra of the molecules are related to their necklaces and are well coherent with the corresponding spectral transformation that is characteristic for the spectra of numerous organic molecules when going from pure oxygen to hydroxylic and carboxylic addends (see for details [11, 12]). Well characteristic GFs related to the O-H stretchings and OH rockings, the former at 2900 cm$^{-1}$ and 3400 cm$^{-1}$, while the latter at 250 cm$^{-1}$ are clearly vivid. The same can be said about C-O, C-O(H), and C=O stretchings at 1500 cm$^{-1}$, 1700 cm$^{-1}$, and 2100 cm$^{-1}$, respectively. C-C(O) bendings at 600 cm$^{-1}$ complete the GFs list related to framed graphene molecules in full accordance with the standard GFs listed in Table 2. These highly terminated molecules have been selected for the study to make the fixation of characteristic GFs related to graphene molecule the most reliable.

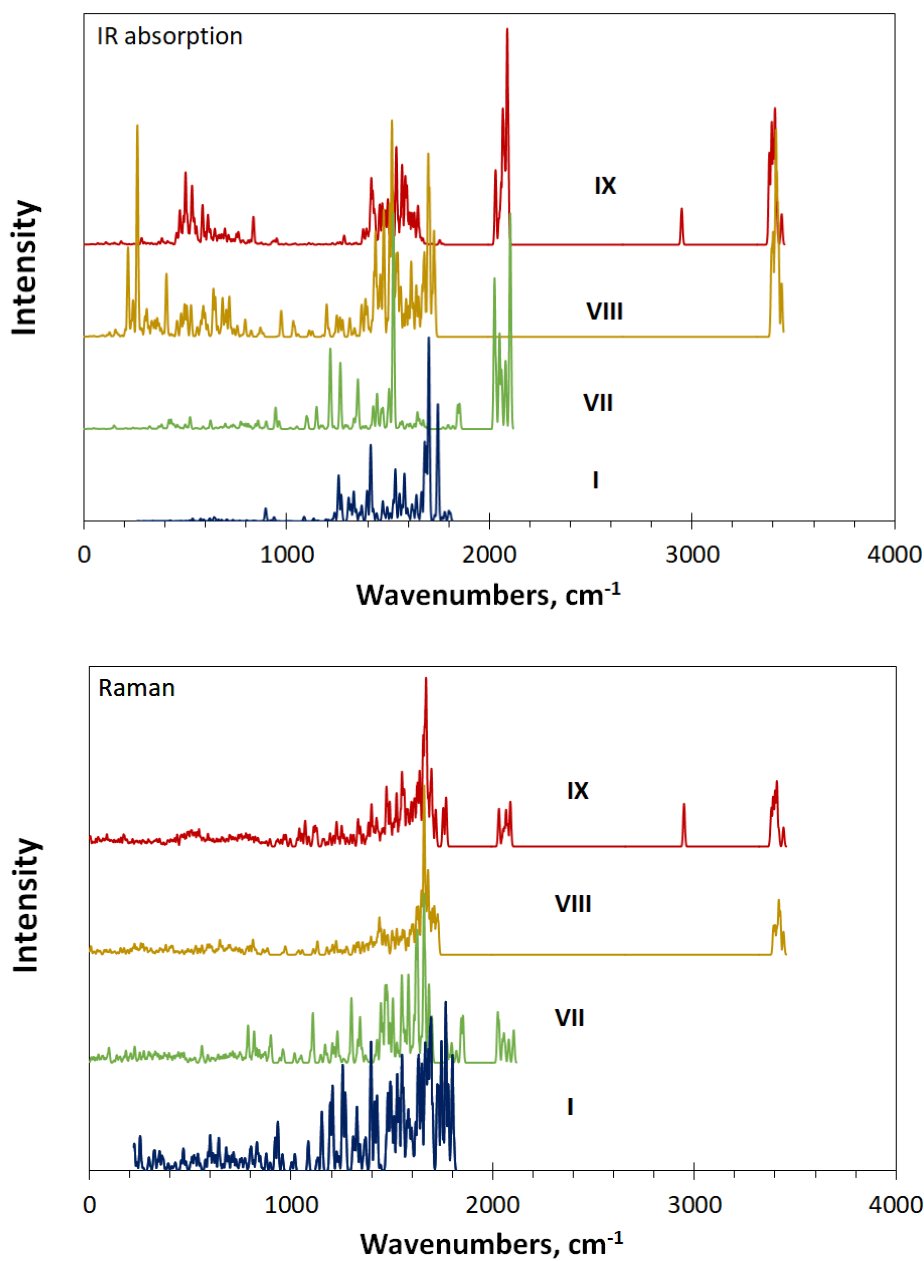

**Figure 7.** Virtual IR absorption and Raman spectra of pristine graphene molecule (**I**, dark blue) with different oxygen necklaces consisting of 22 oxygen atom (**VII**, dark green), 22 hydroxyls (**VIII**, dark yellow), and 20 carboxyls (**IX**, dark red). Stick-bar data are convoluted with Gaussian bandwidth of 10 cm$^{-1}$. Intensities are reported in arbitrary units, normalized per maximum values.

Raman spectra of molecules **VIII** and **IX** are definitely related to the necklaces with the most bright peculiarities related to C-O stretchings at 1600 cm$^{-1}$. The Raman spectrum of molecule **VII** is evidently mixed, involving both C-O and C=C vibrations, which seemingly is caused by a close similarity of atomic mass of carbon and oxygen atoms.

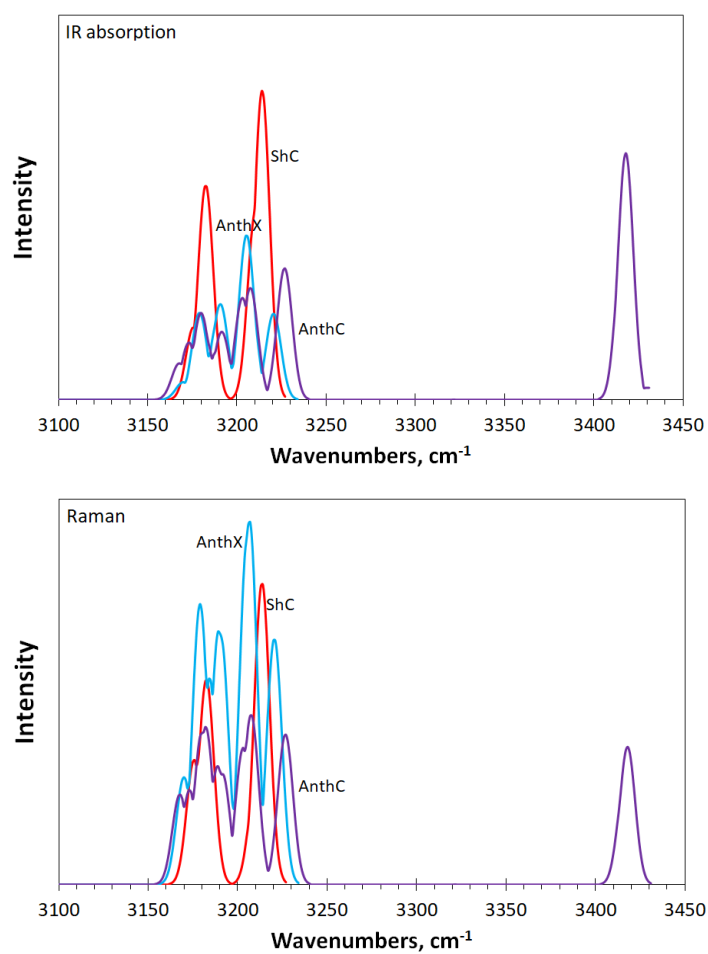

**Figure 8.** Virtual IR absorption and Raman spectra of BSU models of *sp²* amorphous carbons in high-frequency region: shungite carbon (red), anthraxolite (blue), and anthracite (violet). Stick-bar data are convoluted with Gaussian bandwidth of 10 cm$^{-1}$. Intensities are reported in arbitrary units, normalized per maximum values.

## 6. Virtual vibrational spectra of models related to *sp²* amorphous carbons

Diverse experimental studies, recently performed for a selected set of *sp²* amorphous carbons (ACs) [6-8], have allowed suggesting reliable molecular models of their basic structure units (BSUs). The models present nanosize framed graphene molecules constructed on the basis of the pristine (5x5)NGr one. The suggested UHF VVS allows obtaining these molecules virtual vibrational spectra so that a temptation has arisen to compare the latter with previously observed [6, 7]. Four model molecules (**X**, **XI**, **XII**, and **XIII**), equilibrium structure of which is shown in Fig. 1, were selected for the current study. The grounds for the model construction are given elsewhere [6-8]. Model **X** presents anthracite (AnthC), models **XI**, **XII** and **XIII** represent, anthraxolite (AnthX), shungite carbon (ShC), and industrial carbon black (CB632), respectively.

Since C-H and O-H stretchings dominate in the spectra of hydrogenated natural amorphics, thus suppressing the intensity in the middle frequency region and complicating spectra analysis, below high-freq and mid-freq regions will be considered separately. Within each of the part, the spectra will be renormalized with respect to the maximum intensity value. A general picture of the vibrational spectra of the models in the high-freq region is presented in Fig. 8. As seen in the figure, the same set of C-H and O-H stretchings at 3200 cm$^{-1}$ and 3400 cm$^{-1}$, respectively, lay the foundation of both IR and Raman spectra, thus clearly revealing a different disposition of C-H bonds in the necklaces of the studied molecules. Figure 9 presents the set of mid-freq original UHF VVS spectra. As seen in the figure, IR spectra of natural amorphic models from AnthC to ShC are evidently necklace ones thus strongly supporting the main conclusion given elsewhere [6]. The IR spectrum of industrial carbon black seems to be a mixture of C=C and C-O vibrations. A detailed analysis of the spectra is complicated due to small intensity of bands below 1800 cm$^{-1}$, which is caused by a considerable dominating the bands related to the C=O stretchings at 2000 cm$^{-1}$. Aiming at a further comparison of the calculated and empirical DRIFT spectra of the species [6], the intensity of the C=O stretching bands was lowered by a few times to make it relatively similar to the observed in the DRIFT spectra. Certainly, the undertaken measure caused an obvious renormalization of the remain parts of the spectra, the result of which is presented in Fig. 10. The changed plotting allows now a detailed analysis of the IR spectra of the species in the region below 1800 cm$^{-1}$. This analysis should be performed into tight connection with the model atomic structures. Below is a brief overview of the qualitative characteristics of the IR spectra shown in Fig. 10. Let us go from bottom to up.

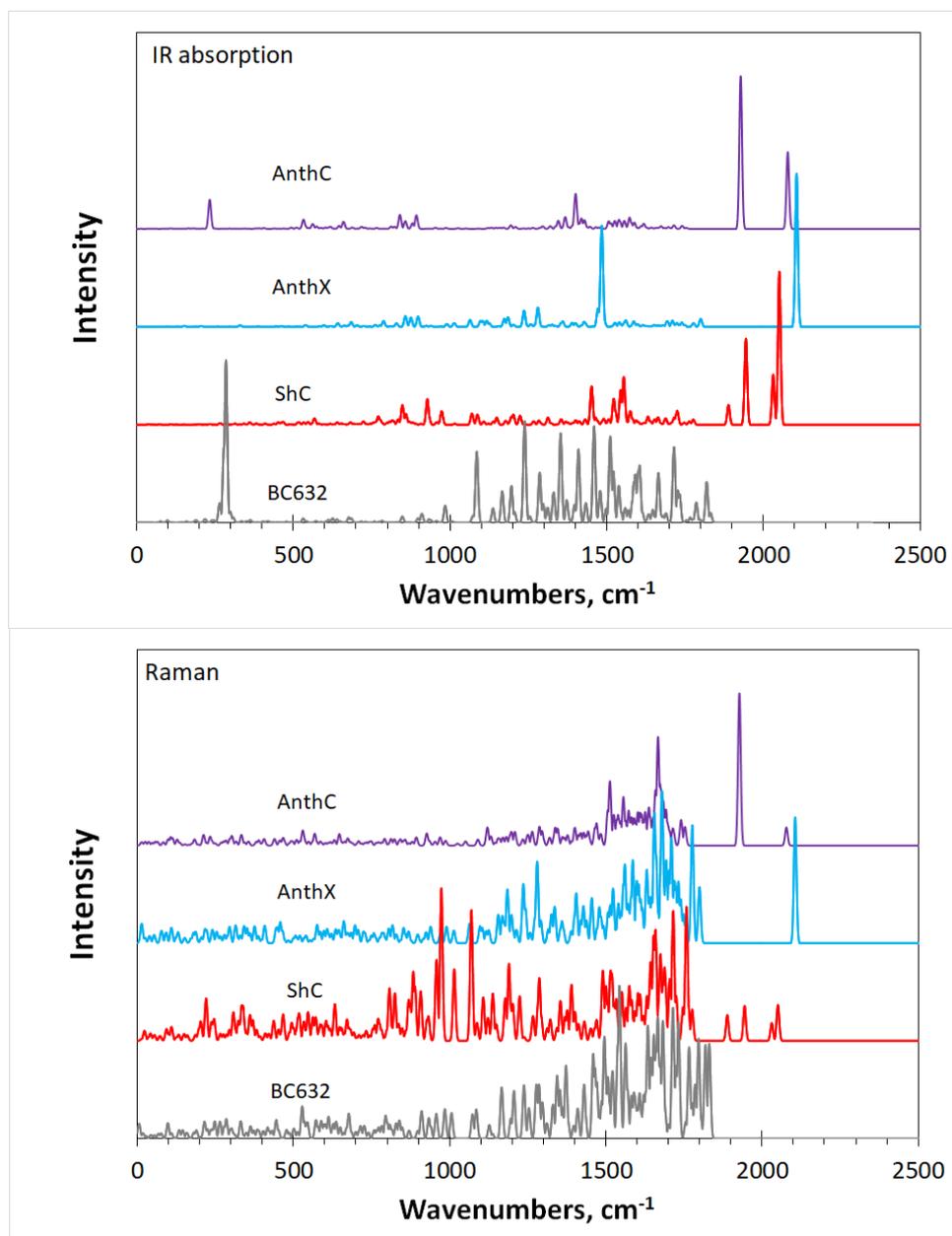

**Figure 9.** Virtual IR absorption and Raman spectra of BSU models of *sp²* amorphous carbons: carbon black (gray), shungite carbon (red), anthraxolite (blue), and anthracite (violet). Stick-bar data are convoluted with Gaussian bandwidth of 10 cm$^{-1}$. Intensities are reported in arbitrary units, normalized per maximum values.

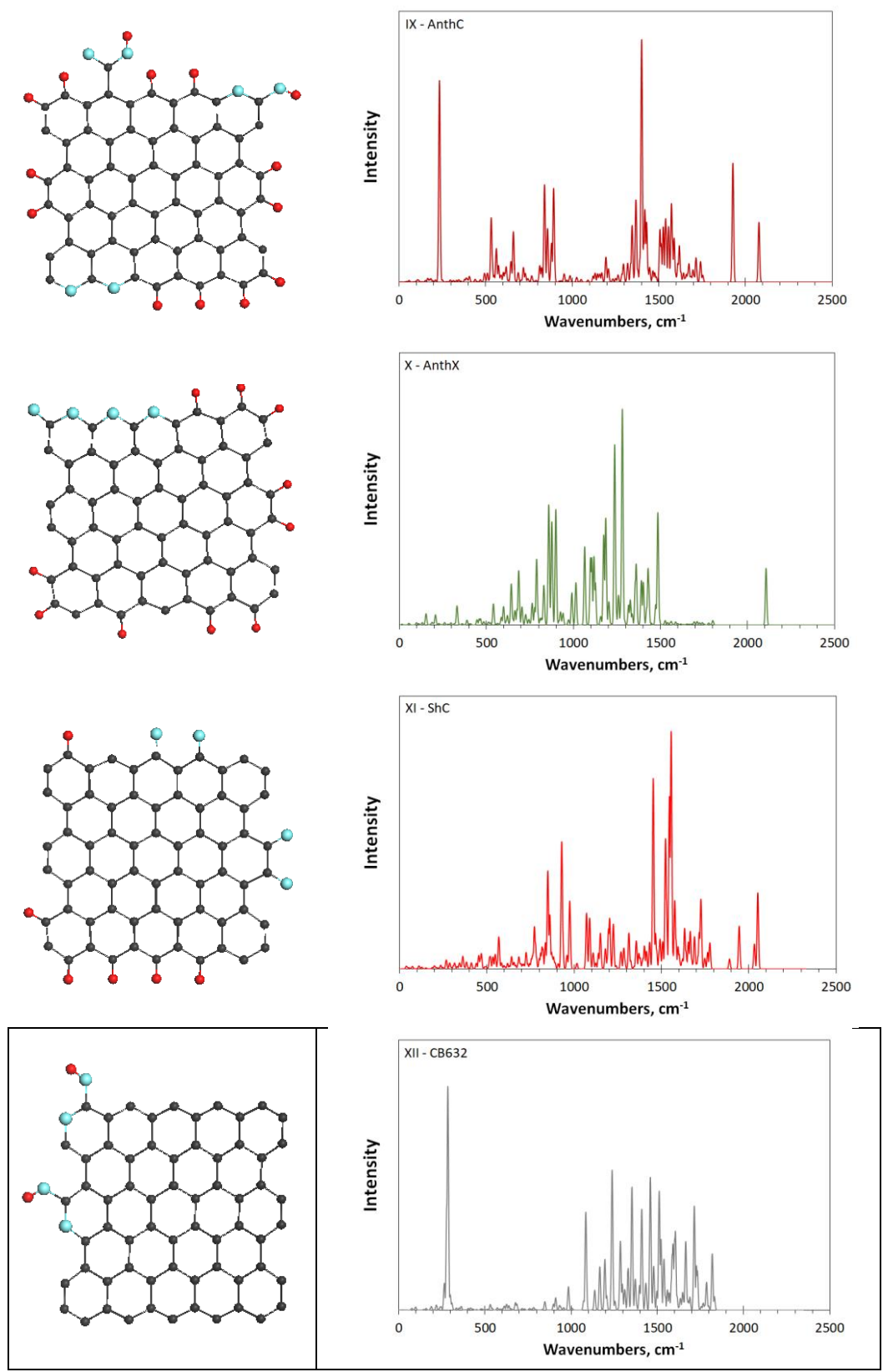

**Figure 10.** Virtual IR absorption spectra of BSU' models of $sp^2$ amorphous carbons, renormalized after scaling in the region of 1900-2100 cm$^{-1}$ (see text): carbon black (gray), shungite carbon (red), anthraxolite (blue), and anthracite (violet). Stick-bar data are convoluted with Gaussian bandwidth of 10 cm$^{-1}$. Intensities are reported in arbitrary units, normalized per maximum values.

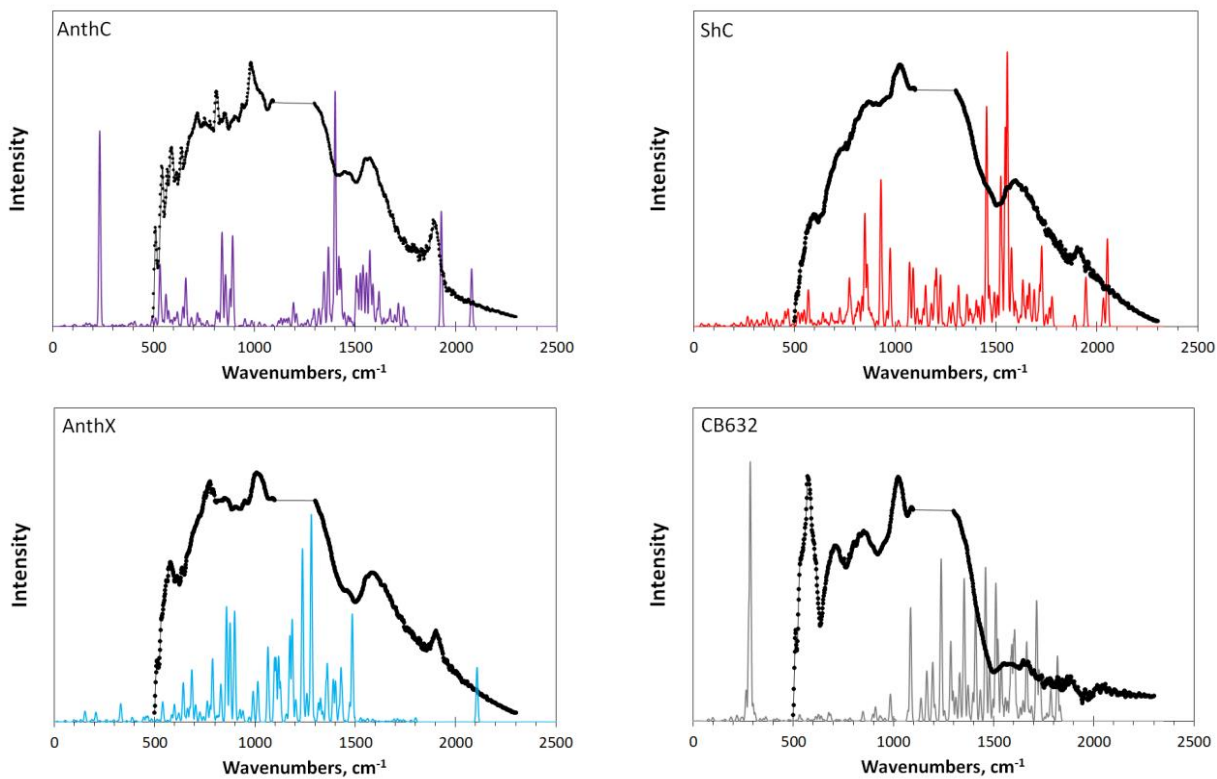

**Figure 11.** Virtual IR absorption spectra of BSU' models of $sp^2$ amorphous carbons, scaled in the region of 1900-2100 cm$^{-1}$ (see text): carbon black (gray), shungite carbon (red), anthraxolite (blue), and anthracite (violet). Stick-bar data are convoluted with Gaussian bandwidth of 10 cm$^{-1}$. Intensities are reported in arbitrary units, normalized per maximum values.

*Molecule **XIII** – CB632*. The presence of only four oxygen atoms and two hydrogen atoms, which is only 8.6 at% of the total molecular composition, radically changes the IR spectrum in comparison with the spectrum of the pristine (5x5)NGr molecule **I** shown in Fig. 5. The intense spike at 270 cm$^{-1}$ corresponds to OH torsions. The spectrum in at the region from 1000 cm$^{-1}$ t0 1800 cm$^{-1}$ is a peculiar mixture of C-O and C=C bendings and stretchings and needs a further studying.

*Molecule **XII** – ShC*. The necklace atomic content constitutes 13 at % and the spectrum has an obvious necklace character. Carbonyl units provide two intense bands at 2000 cm-1, apparently distinguishing their location on the zigzag and armchair edge atoms. The simultaneous presence of hydrogen and oxygen in the molecule necklace results in a its bizarre spectrum in the 700-1800 cm$^{-1}$ region, mixing C-H bendings and stretchings with those stimulated by C=O bonds.

Similar discussions can be proposed for *Molecules **XI** – AnthX* and ***X**- Anthc*. Finally, Fig. 11 presents the comparison between virtual and experimental data for the amorphics' models. Since upshifting of the UHF VVS spectra is different for different spectral region, to partially compensate it all experimental are shifted up by 100 cm$^{-1}$ and additional shifting of 200 cm$^{-1}$ is applied at the region above 1200 cm$^{-1}$. It should be noted that experimental sample were powdered solids, consisting from multi-layered stacks of nanosize (about 1,5 nm) graphene molecules similar to those shown in the left row in Fig. 10, aggregated in globules and large agglomerates. As was seen in Fig. 2, the layering significantly influences the spectra image due to which the picture exhibited in Fig. 11 can be considered as a good fitting since the conclusion about particular GFs participating in the spectra formation correlates well with that made when scrupulously analyzed empirical spectra [6, 7]. Certainly, a lot of work is to be done for the exhausted analysis and a complete understanding the spectra structure to be obtained. However, even this first attempt to use for the goal *in silico* spectroscopy looks quite promising.

## 7. Conclusion

The results presented in this article show that *in silico* spectroscopy can become an effective method for studying the vibrational spectra of large molecules and, particularly, radicals. The first application of the UHF VVS spectrometer to graphene molecules made it possible to establish hitherto unknown features of the formation of their vibrational spectra. A deep analysis of these features has yet to be done, but it is already obvious that we are expecting the discovery of new sides of the special electronic structure of graphene. The obtained data open the way to a new wide area of research in this area, full of new unexpected discoveries.

**Acknowledgements**

The authors are thankful to Ye. A. Golubev for fruitful discussions and to V. V. Kim for assisting with a software in the spectra treatment. The publication has been prepared with the support of the "RUDN University Program 5-100".